# THE RECURRENT NOVA T CRB
# DID *NOT* ERUPT IN THE YEAR 1842

*By Bradley E. Schaefer*
*Department of Physics and Astronomy, Louisiana State University*

The recurrent nova T CrB was one of the first well observed nova eruptions in 1866, and 80 years later it erupted again in 1946. Just after the 1866 eruption, Sir John Herschel reported to the *Monthly Notices* that he had seen the same star in his naked-eye charting of the sky on 1842 June 9, implying that there was a prior eruption 24 years earlier. Unfortunately, the chart in the *Monthly Notices* was ambiguous and misleading, so it has long been unclear whether T CrB did indeed have an eruption in 1842. To resolve this, I have searched the various archives with Herschel material, and have found his original correspondence. In one letter from 1866 to William Huggins, Herschel enclosed his own copy of his original observations, and with this all the ambiguities are resolved. It turns out that Herschel's indicated star was at the same position as a steady background star (BD+25°3020, V=7.06, G8V) and not that of T CrB, and Herschel regularly was seeing stars as faint as V=7.5 mag because he was using an opera glass. With this, there is no evidence for a T CrB eruption in 1842.

*Introduction*

A recurrent nova (RN) is a close binary star where one star spills matter, through its Roche lobe, onto a white dwarf, where the material accumulates until a runaway thermonuclear explosion is triggered. RNe are distinct from ordinary novae primarily due to their recurrence time scale being less than a century or so[1]. To achieve this fast recurrence, the white dwarf must be near the Chandrasekhar mass and the accretion rate must be relatively high. These conditions are exactly as expected to lead to the white dwarf gaining mass until it would explode as a Type Ia supernova. As such, RNe are one of the best candidates for being a supernova progenitor. The identity of the progenitor has been one of the keystone problems in stellar evolution for over three decades, and since the late 1990's has become of even higher importance because the progenitor must be known so as to calculate the evolution of Type Ia supernovae to high redshift, with this now being the dominant uncertainty in the several very large-scale programs of precision cosmology.

One of the primary uncertainties for whether RNe are progenitors are their recurrence time scales, as this determines whether their white dwarfs are gaining or losing mass over each eruption cycle as well as the number of RNe in our Milky Way and hence whether the RNe death rate matches the Type Ia supernova rate[2,3]. For this reason, I have spent roughly a man-year examining plate archives for prior eruptions of known RNe and looking for second eruptions on candidate RNe, with the result of discovering one new RN, 6 RN eruptions, and predicted/confirmed one more[1,4]. A conclusion from this work is that the average recurrence time scale has been shortened by roughly a factor of three.

T Coronae Borealis (T CrB) is one of the ten known RNe in our own galaxy, with well-recorded eruptions in 1866 and 1946. T CrB is normally in quiescence at V=9.8 mag, while it peaked during its eruptions at V=2.5, making it one of the brightest novae in the last two centuries. Like roughly half the RNe, the companion star is a red giant star, with an orbital period of 228 days. T CrB has been monitored frequently enough from 1866 to 2012 (including roughly hourly to weekly from 1890 to 2012) that we know that an eruption cannot have been missed, because even an eruption around the time of solar conjunction would have been spotted after it comes out in the morning sky[1]. With this, the recurrence timescale is apparently 80 years, with the next eruption expected around the year 2026.

*Herschel's Monthly Notices Article in 1866*

Soon after the 1866 eruption of T CrB, Sir John F. W. Herschel published a short article in the *Monthly Notices*[5] pointing out that he had seen the same star on 1842 June 9. Back at the time, the recognition of novae as a class was unknown (indeed, T CrB was only the third nova seen in the prior 80 years), so the implications of his observation were not known. With the modern understanding of RNe, we see this record as possibly being of a prior nova event. Suddenly, the average recurrence timescale for T CrB would change from 80 years to 52 years, (1946-1842)/2, with the implication that T CrB should go off anytime soon. Further, the inter--eruption interval would vary by at least a factor of 3.3 (from 24 years to 80 years), with such a variation being inexplicable. This would be substantially different from the other RNe, where the recurrence time--scales change by 20% for U Sco and 42% for RS Oph[1]. (The case of T Pyx is unique and represents a greatly different case[6].) So there is good modern astrophysics utility in knowing whether T CrB erupted in 1842.

In 1939, Dean McLaughlin examined the evidence for the 1842 eruption[7]. He had access only to the *Monthly Notices* report. McLaughlin concluded that Herschel had not actually seen T CrB in 1842, but had rather merely recorded the nearby star BD+25°3020 (HD 144287). The sole basis for this conclusion was that the position of one of the points in the *Monthly Notices* chart is roughly one degree from the position of T CrB and is apparently coincident with the background star. One degree (i.e., two Full Moon diameters) is a substantial distance on the sky, so he suggested that this meant that the 1842 star recorded could not be T CrB.

This conclusion has three severe problems. First, it is unclear which of two symbols on the *Monthly Notices* chart refers to the star observed by Herschel in 1842. The star labeled as 6' is near the BD star, while the asterisk is near the position of T CrB, with it being unclear what, if anything, was added by the editor. Second, the *Monthly Notices* chart has the faintest recorded star with V=5.82, so it seems that Herschel could not have seen BD+25°3020 at V=7.06. This star is of spectral type G8V and has never been seen to vary in brightness. Third, not only did Herschel apparently not see below sixth magnitude, but it is very unlikely for anyone to see to V=7.06 under any conditions. All pre-telescopic star catalogs have the 50% limiting magnitude of near fifth magnitude, and I have extensively tested large numbers of people under dark skies with the same result. (Star detection near the limit is a probabilistic task, where the detection probability falls off from near unity to near zero typically over a range of one magnitude, with the probability being 50% in the middle of this range.) This is not to say that it is impossible to see to V=7.06 under optimal conditions, as, for example, I have recorded Stephen O'Meara (one of the premier visual observers for the last century) as seeing stars as faint as V=8.2 from the top of Mauna Kea[8]. While few humans can see so faint, it is very unlikely that Herschel would have recorded the BD star without telescopic aid even on a clear moonless night from a near-sea-level site in humid England.

So the real status of the 1842 eruption is unknown due to the ambiguities in the published chart. The only way to resolve this is to find the original records from 1866 or 1842.

*Herschel's Correspondence and Diary*

Sir John Herschel was amongst the greatest visual observers of his century. In 1838, he arrived back from four years in South Africa mapping out the southern skies, completing the task started by his father of cataloging the nebulae and double stars over the whole sky. In April of 1840, he moved to Kent, to a house called Collingwood, where he continued to make a few observations with small telescopes, and to analyze and publish his South African observations[9].

At least from 1841 January to 1842 June, Herschel made naked eye charts of the entire northern skies showing all visible stars[10-12]. This was an extension of a program he began in South Africa to map the skies and assign magnitudes to all stars. He had divided up the sky into 738 (mostly) triangles[12], with T CrB being inside the triangle formed by α CrB, β Her, and ζ Her[11]. I have not found anywhere where this data was published, but Herschel certainly kept his charts at least until 1866. In a letter to the Reverend C.

Pritchard dated 1867 April 10, Herschel says "Thanks for the care of binding up my Star allineations."[13], so apparently the star charts became bound around 1867. Herschel tells how his star charts were donated to the Royal Astronomical Society (RAS), and he provides some details on their construction[14].

In a letter to Francis Baily dated 1841 January 26, Herschel states "You will receive by Wednesday's Coach a roll containing a circumpolar projection of the Southern Stars as far as 70° SPQ [south polar distance] in which all the stars of Magnitudes 1.2.3.4.5 are laid down as they really appear to the naked eye and many of the magnitudes (5.6-) [i.e., stars of fifth magnitude, sixth magnitude, and fainter] When I say all I mean that I have examined seriatim [in series] every triangle of stars forming a network over the whole - about 360 triangles - carefully with naked [eye] occasionally (and usually at last) aided by an opera glass laying down in each, on great charts pricked off from Bode's maps, all the discernable stars down to 7m[ag]"[10]. What this means is that Herschel was only picking out the brighter stars (as faint as magnitude 5) with the unaided eye, but that the fainter stars were observed with a small telescope (an opera glass). This is confirmed by page headers from 1841 in Herschel's notebooks[15] summarizing magnitudes from prior star triangles, which state "Magnitudes for Naked Eye and Opera Glass as read off on the Working Charts". Unfortunately, I can find no specific information on the properties of the opera glass, but historical examples[16] suggest that it was two classical Galilean telescopes with perhaps 2 centimeter aperture mounted for binocular vision and having a fairly small (perhaps 3°) field of view. Now, we have an explanation for why Herschel was able to report stars much fainter than almost all other humans, and it was simply because he was using an opera glass.

In 1866, Herschel was one of the pre-eminent scientists in the world, having made strong advances in astronomy, physics, botany, chemistry, and photography. He was frequent in consultation and communication with astronomers around the world. In 1866 May 14, the eruption of T CrB was discovered by John Birmingham in Ireland, and Herschel was quickly notified by William Huggins. Herschel then looked at his 1842 chart and found a star at nearly the right position. In a letter dated 1866 May 19, Herschel tells Huggins about his old "Star-triangles" and the star at the position of the nova (labeled 6') as observed on 1842 June 9[17]. Critically, this letter had attached a copy that Herschel made of his original 1842 chart, with this having been made by pricking the stars from the overlain original (see Figure 1). In a letter to the Astronomer Royal George B. Airy dated May 25, he writes about his star (which he says has magnitude 6⅓) and explicitly that its position is within the "limits of error" of T CrB. Airy replied on May 26 that Herschel should place a letter into the *Monthly Notices*[18]. On May 29, Herschel complies with a letter for publication, that must have contained a chart[5], but I have found no evidence of this letter or chart having survived. On May 29, Edward James Stone (then Chief Assistant at the Royal Greenwich Observatory and later President of the RAS) acknowledges receipt of Herschel's contribution, including the chart for publication, and provides some short discussion of spectral lines visually seen by himself in the previous night[19]. Stone enclosed his version of Herschel's chart, very similar to that which appeared in the *Monthly Notices*, except that Herschel's 1842 star is explicitly identified as being the one labeled 6' that is at the position of BD+25°3020 (see Figure 2), while this is distinctly separated from T CrB (labeled "variable"). Stone's copy shows that he did not reproduce most of the faint stars, hence making for the impression that the limiting magnitude was not even as deep as 6. The same chart was then published in the *Monthly Notices*, except that the star labeled "Hershel" was called 6' and the star at the position of T CrB was no longer labeled, hence leading to the ambiguity of the printed chart. Thus, Stone's simplifications of Herschel's chart made for all the problems in knowing whether he had seen T CrB in 1842.

 The only previously recorded magnitude for T CrB was 9.5 as given by Friedrich W. A. Argelander in the BD catalog. On 1866 October 29, Argelander[20] wrote to Herschel asking whether his 1842 star could be the background star now called BD+25°3020. On 1866 November 6, Herschel[21] replied "My star 6' m[ag] was laid down without that particular precision as to allineations and which would of course have been given had I had any suspicion of its being a remarkable object and though its place does not agree very well with the Variable, I do not think the difference more then <u>might</u> have occurred." The underlining is Herschel's. So the positional offset between Herschel's 1842 star and the position of T CrB (as noted by Argelander and McLaughlin) is within his normal uncertainty, although the words and the underlining suggest that the positional difference is substantially larger than is usual.

 For the modern astrophysics question, the key item is Herschel's copy of his 1842 chart (see Fig. I) that accompanied the May 19 letter to Huggins. This chart answers all the questions raised by the *Monthly Notices* chart. First, the 1842 star is clearly identified with a distinct arrow as being the one labeled 6', and this is clearly at the position of BD+25°3020 and roughly a degree from the position of T CrB. (In Figure 1, the star is close to the center and pointed at with an arrow.) Second, the chart shows many stars going very faint, including the memorable comparison star next to R CrB (HD 141352 at V=7.48), plus HD 139608 (V=6.88), HD 146604 (V=6.55), HD 145457 (V=6.59), and HD 145976 (V=6.48). The brightest star that is certainly missing from within the triangular chart area and within 5° of T CrB is HD 142053 (V=7.46). So Herschel's real limiting magnitude is roughly 7.5. Third, this deep limit is possible simply because he was using an opera glass. With this, we see that Herschel could easily see BD+25°3020. Indeed, given this real limiting magnitude, the BD star should appear on the 1842 chart, and if the arrowed star is not the BD star then we have the dilemma of asking why that one star was not recorded. With this, Herschel's 1842 chart makes sense only if the star recorded is BD+25°3020, while T CrB was not visible. As such, all evidence for an 1842 eruption of T CrB vanishes.

*Herschel's Original 1842 Chart*

   The 1866 copy made by Herschel by pricking through paper of his original 1842 chart has answered all the astrophysical questions.  Nevertheless, there is a historical imperative to examine the original source if at all possible.  To this end, I have sought long and hard for the original 1842 chart.  I have flown to Austin Texas and examined all astronomy documents in the large collection of John Herschel material at the Harry Ransom Center.  The complete collection of all Herschel material now archived by the RAS has been digitally photographed and placed on 17 CDs, and I have gone through all the pages of John Herschel's material.  All Herschel letters in archives worldwide have been exhaustively cataloged, with this available in book form[22] and at the Adler Planetarium web site[23].  The ~10,000 Herschel letters in the Royal Society archives have been copied and are all available in microfilm[24].  I have searched the archives at St. John's College, Cambridge[25], the Science Museum of London, and the Royal Greenwich Observatory, Cambridge.  I have consulted with the current head of the Herschel family (Mr. John Herschel-Shorland), the leading Herschel historians (Drs. Michael Hoskin and Michael Crowe), and the leading astronomers who have worked on the historical T CrB (Drs. Ronald Webbink and Brian Warner).  With Herschel telling us that his charts were donated to the RAS[14], the RAS archives are the most likely place to find the 1842 chart.  But the entire collection has been catalogued[26] and photographed (with the CDs available for purchase from the RAS), with no chart from 1842.  RAS librarian, Peter Hingley, authoritatively tells me that there is certainly no Herschel manuscript or document that was not included into the CD version.  The original 1842 June 9 chart of Herschel is not to be found.
   I am convinced that the 1842 chart still exists, because the chart was saved from at least 1842 to 1866, because John Herschel showed strong interest in preserving papers from his father and himself, and because Herschel's children were actively involved in preserving his material.  Apparently it was bound with other original charts and donated to the RAS and they would never get rid of such a document, but it is not now to be found in their complete collection as recorded on CDs.
   Despite the 1842 chart not being found, we have a fair copy in Herschel's own hand from 1866, and this resolves the astrophysics question.  In 1842, Herschel was recording stars regularly as faint as V=7.5 with the aid of his opera glass, and his 1842 chart simply showed the ordinary star BD+25°3020.  T CrB did not erupt in 1842.

*Acknowledgments*

I thank Brian Warner, Ronald Webbink, Michael Hoskin, Michael Crowe, Steve Ruskin, Marv Bolt, Stephen Case, Peter Hingley, and John Herschel-Shorland. I also thank the archives and staff at the Harry Ransom Center, the Royal Astronomical Society, the Royal Society, Adler Planetarium, and the Science Museum of London. This research is supported by the National Science Foundation, through grant AST-1109420.

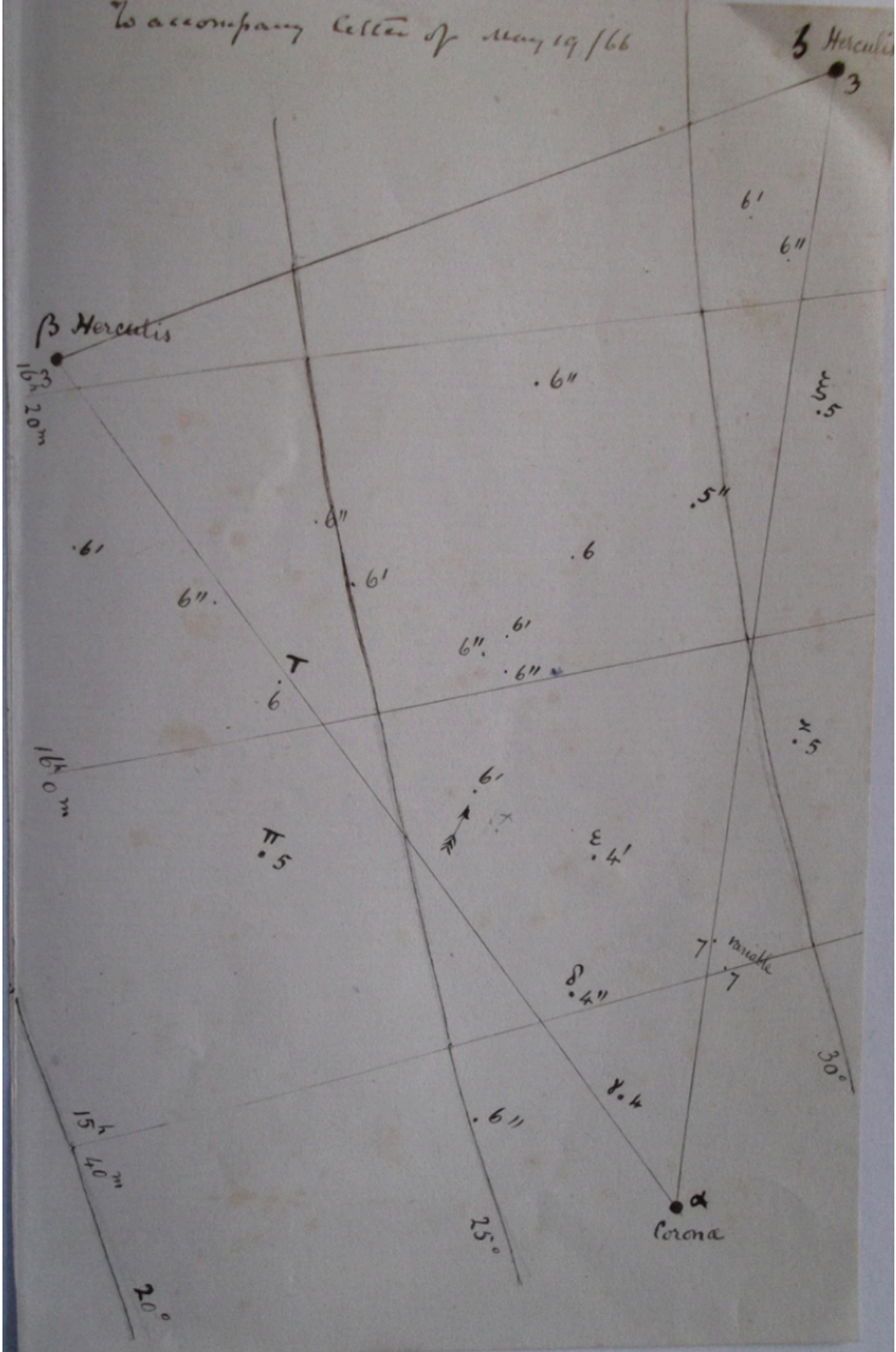

Fig. I

This chart is from a letter sent by John Herschel[15] to William Huggins on 19 May 1866 (© Royal Society). The basic triangle stretches from α CrB (near the bottom) to β Her (near the top left) to ζ Her (in the upper right corner). In addition to the lines giving the boundaries of the triangle, Herschel also places lines for a coordinate grid with right ascensions from 15h 40m to 16h 20m as well as declinations from +20° to +30° (for the equinox of 1860). The star that Herschel claimed was T CrB is near the center and labeled with an arrow and the magnitude notation 6'. Importantly, this star is right at the position of BD+25°3020 and about one degree away from the position of T CrB. The positional accuracy of this copy (relative to the 1842 chart) is ensured by Hershel having `pricked' a needle through the original chart onto a lower sheet of paper. Just below the middle on the right are a pair of stars, R CrB (labeled "variable") and its nearby comparison star HD 141352 at V=7.48. So Herschel was certainly seeing faint enough to pick up BD+25°3020. The brightest star that is certainly missing is V=7.46. So Herschel's real limit in this chart is close to 7.5 mag, with this being reasonable because we know from his correspondence and notebooks that he made his star chart with the help of an opera glass. So, Herschel could easily see BD+25°3020, this star should appear on his chart, and the charted star is at the position of BD+25°3020, so the evidence shows that this charted star is not T CrB in eruption in 1842.

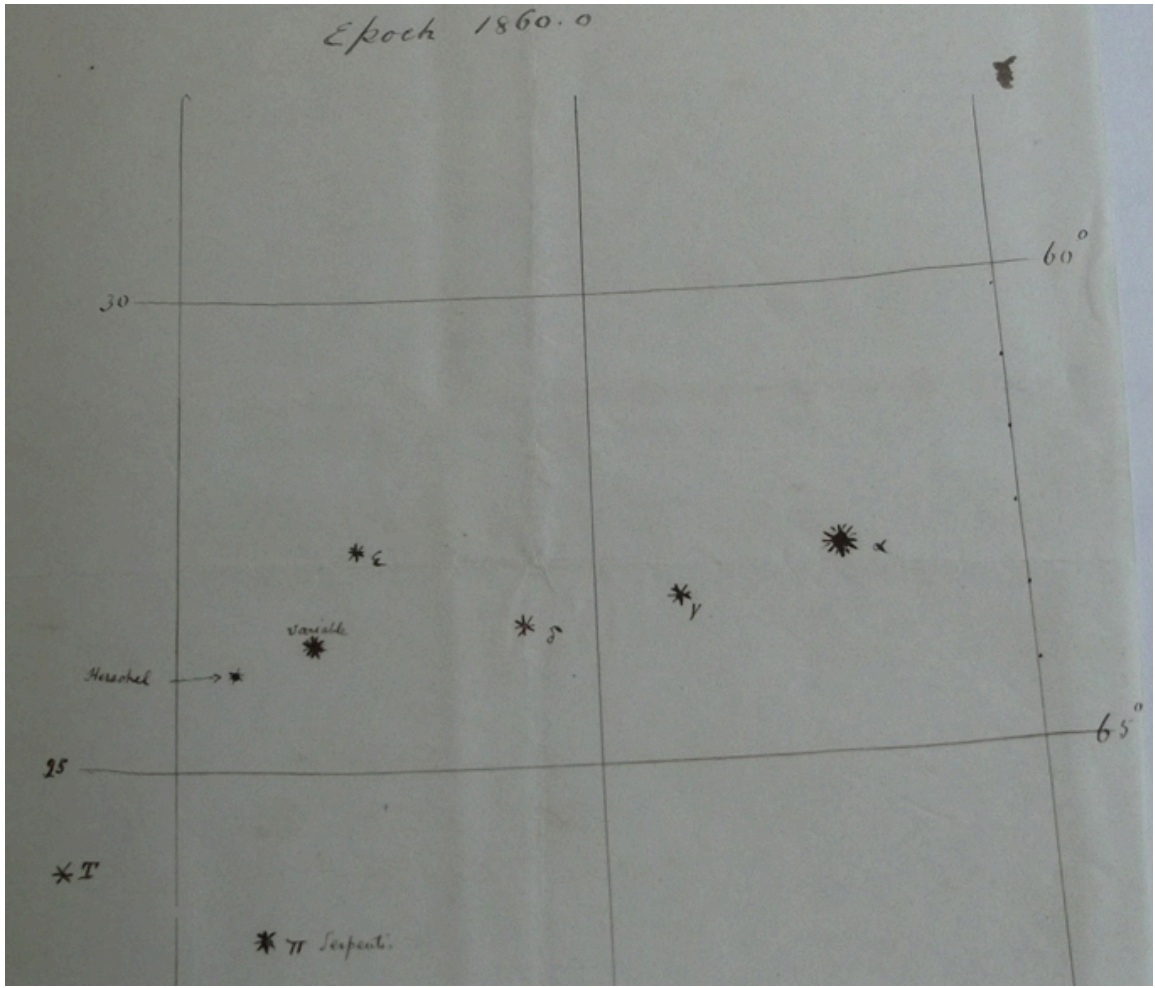

Fig. II

This chart is from a letter sent by E. J. Stone[19] to John Herschel on 29 May 1866 (© Royal Society).  Herschel had just submitted his short paper to the *Monthly Notices*, including the chart similar to that in Figure 1, and Stone had replied back in his capacity as Chief Assistant to Airy while including this chart for Herschel's inspection.  Critically, this chart clearly shows that the position of Herschel's 1842 star (labeled "Herschel") is substantially different from the position of T CrB (labeled "variable").  Also, Stone's simplification of Herschel's chart has deleted all the faint stars, thus giving evidence that the limiting magnitude of Herschel was around fifth magnitude, with the erroneous implication that Herschel could not possibly have seen BD+25°3020 at V=7.06.  Stone further compounded the confusion when the chart published in the *Monthly Notices* changed the labels, from "Herschel" to 6' and from "variable" to blank, thus making it unclear as to the observed position of the star in 1842.